\documentclass[%
 reprint,
 superscriptaddress,
bibnotes,
amsmath,amssymb,
aps, 
pra,
]{revtex4-2}
\usepackage{graphicx} 
\usepackage[caption=false,subrefformat=parens,labelformat=parens]{subfig}
\usepackage{dcolumn}
\usepackage{bm}
\usepackage{hyperref}
\hypersetup{
    colorlinks=true,
    citecolor=blue,
    linkcolor=blue,
    filecolor=blue,  
    urlcolor=blue,
    pdftitle={A Framework for Achieving Something from Nothing},
    pdfpagemode=FullScreen,
    }
\usepackage[mathlines]{lineno}
\usepackage{color}
\usepackage{physics}
\usepackage{dsfont}
\usepackage{mathtools}
\usepackage{lipsum}
\usepackage[top=2cm, bottom=2cm, left=1.5cm, right=1.5cm]{geometry}
\usepackage{hyperref}
\usepackage[dvipsnames]{xcolor} 

\usepackage{orcidlink}

\usepackage[colorinlistoftodos]{todonotes}
\usepackage{xargs}
\setuptodonotes{size=\tiny,linecolor=Plum,backgroundcolor=Plum!25,bordercolor=Plum}

\newcommand{\ad}{a^{\dagger}}
\newcommand{\bd}{b^{\dagger}}
\newcommand{\rme}{{e}}
\newcommand{\rmi}{{i}}
\newcommand{\rmd}{{d}}

\newcommand{\VV}{\mathbf{V}}

\begin{document}

\preprint{APS/123-QED}

\title{Something from Nothing: \\ A Theoretical Framework for Enhancing or Enabling Cooling of a Mechanical Resonator via the anti-Stokes or Stokes Interaction and Zero-Photon Detection }

\begin{abstract}
We develop a theoretical framework to describe how zero-photon detection may be utilized to enhance laser cooling via the anti-Stokes interaction and, somewhat surprisingly, enable cooling via the Stokes interaction commonly associated with heating.
Our description includes both pulsed and continuous measurements as well as optical detection efficiency and open-system dynamics.
For both cases, we discuss how the cooling depends on the system parameters such as detection efficiency and optomechanical cooperativity, and we study the continuous-measurement-induced dynamics, contrasting to single-photon detection events.
For the Stokes case, we explore the interplay between cooling and heating via optomechanical parametric amplification, and we find the efficiency required to cool a mechanical oscillator via zero-photon detection. 
This work serves as a companion article to the recent experiment~[\href{https://arxiv.org/abs/2408.01734}{E. A. Cryer-Jenkins, K. D. Major, \emph{et al.,} arXiv:2408.01734 (2024)}], which demonstrated enhanced laser cooling of a mechanical oscillator via zero-photon detection on the anti-Stokes signal. The framework developed here provides new approaches for cooling mechanical resonators that can be applied to a wide range of areas including nonclassical state preparation, quantum thermodynamics, and avoiding the often unwanted heating effects of parametric amplification. 
\end{abstract}

\author{Jack Clarke\,\orcidlink{0000-0001-8055-449X}}
\thanks{These authors contributed equally to this work and are listed alphabetically.}
\affiliation{Quantum Measurement Lab, Blackett Laboratory, Imperial College London, London SW7 2BW, United Kingdom}
\author{Evan A.~Cryer-Jenkins\,\orcidlink{0000-0003-2549-0280}}
\thanks{These authors contributed equally to this work and are listed alphabetically.}
\affiliation{Quantum Measurement Lab, Blackett Laboratory, Imperial College London, London SW7 2BW, United Kingdom}
\author{Arjun Gupta\,\orcidlink{0009-0009-2994-9125}}
\thanks{These authors contributed equally to this work and are listed alphabetically.}
\affiliation{Quantum Measurement Lab, Blackett Laboratory, Imperial College London, London SW7 2BW, United Kingdom}
\author{Kyle D.~Major\,\orcidlink{0000-0002-3268-6946}}
\thanks{These authors contributed equally to this work and are listed alphabetically.}
\affiliation{Quantum Measurement Lab, Blackett Laboratory, Imperial College London, London SW7 2BW, United Kingdom}
\author{Jinglei Zhang}
\affiliation{Institute for Quantum Computing, University of Waterloo, Waterloo, Ontario, N2L 3G1, Canada}
\affiliation{Department of Physics \& Astronomy, University of Waterloo, Waterloo, Ontario, N2L 3G1, Canada}
\author{Georg Enzian\,\orcidlink{0000-0002-2603-2874}}
\affiliation{Quantum Measurement Lab, Blackett Laboratory, Imperial College London, London SW7 2BW, United Kingdom}
\affiliation{Clarendon Laboratory, Department of Physics, University of Oxford, OX1 3PU, United Kingdom}
\author{Magdalena Szczykulska\,\orcidlink{0000-0002-5820-7093}}
\affiliation{Clarendon Laboratory, Department of Physics, University of Oxford, OX1 3PU, United Kingdom}
\author{Anthony C.~Leung\,\orcidlink{0000-0001-9420-292X}}
\affiliation{Quantum Measurement Lab, Blackett Laboratory, Imperial College London, London SW7 2BW, United Kingdom}
\author{Harsh Rathee\,\orcidlink{0009-0006-7762-8543}}
\affiliation{Quantum Measurement Lab, Blackett Laboratory, Imperial College London, London SW7 2BW, United Kingdom}
\author{Andreas {\O}.~Svela\,\orcidlink{0000-0002-3534-3324}}
\affiliation{Quantum Measurement Lab, Blackett Laboratory, Imperial College London, London SW7 2BW, United Kingdom}
\author{Anthony K.~C.~Tan\,\orcidlink{0000-0002-1324-4376}}
\affiliation{Quantum Measurement Lab, Blackett Laboratory, Imperial College London, London SW7 2BW, United Kingdom}
\author{Almut Beige\,\orcidlink{0000-0001-7230-4220}}
\affiliation{The School of Physics and Astronomy, University of Leeds, Leeds LS2 9JT, United Kingdom}
\author{Klaus M{\o}lmer\,\orcidlink{0000-0002-2372-869X}}
\affiliation{Niels Bohr Institute, University of Copenhagen, Blegdamsvej 17, 2100 Copenhagen, Denmark}
\author{Michael R.~Vanner\,\orcidlink{0000-0001-9816-5994}}%
\email{m.vanner@imperial.ac.uk}
\homepage{www.qmeas.net}
\affiliation{Quantum Measurement Lab, Blackett Laboratory, Imperial College London, London SW7 2BW, United Kingdom}

\maketitle

\date{\today} 

\section{Introduction}

In quantum mechanics, the act of measurement fundamentally affects the state of a system unlocking capabilities beyond what is classically achievable across a wide range of areas including computing~\cite{briegel2009measurement}, metrology~\cite{giovannetti2011advances}, and communication~\cite{gisin2007quantum}. Over the past few decades, measurement strategies employing single-photon detection have been a vital component in the development of many facets of experimental quantum optics. In particular, single-photon sources via spontaneous parametric down-conversion~\cite{hong1986experimental,eisaman2011invited} and the teleportation of optical qubits~\cite{bouwmeester1997experimental,boschi1998experimental,pirandola2015advances}. Single-photon detection has also proved to be a crucial resource to engineer of non-Gaussian quantum states with applications in quantum sensing~\cite{lachman2022quantum} and investigations of fundamental physics~\cite{zavatta2009experimental}. In purely optical systems, single-photon measurements, together with the beamsplitter and two-mode squeezing interactions, have been used to prepare nonclassical quantum states of light via photon subtraction~\cite{dakna1997generating,ourjoumtsev2006generating,neergaardNielsen2006} and addition operations~\cite{agarwal1991nonclassical,zavatta2004quantum,zavatta2007experimental}, respectively. Notably, in the field of optomechanics, driving the anti-Stokes and Stokes scattering processes naturally enables the same form of beamsplitter and two-mode squeezing interactions between an optical field and a mechanical resonator~\cite{meystre2013short,aspelmeyer2014cavity}. Using these interactions, optomechanics now provides a platform to achieve single-phonon addition and subtraction operations~\cite{vanner2013quantum}. Experimentally, such addition and subtraction operations have been performed on thermal mechanical states, which increases the mean mechanical occupation~\cite{Enzian2021}, and multi-phonon operations to generate mechanical non-Gaussianity has also been performed~\cite{Enzian2021_2,Patel2021}. Additionally, higher-order phonon coherences~\cite{patil2022measuring} and nonclassical correlations are now being investigated~\cite{lee2012macroscopic,riedinger2016non,galinskiy2023non}.

In the experiments described above, single-photon detection is employed, however, the absence of a photon at the photon counter also provides valuable information that can be utilized in a variety of applications. In particular, the detection of zero photons enables probabilistic noiseless attenuation and amplification for quantum communication~\cite{xiang2010heralded,mivcuda2012noiseless,gagatsos2014heralded,brewster2017noiseless,ye2019improvement}, optical state reconstruction~\cite{banaszek1996direct,wallentowitz1996unbalanced,kim1997quasiprobability,rossi2004photon,zambra2005experimental}, Gaussification protocols for entanglement distillation~\cite{browne2003driving,campbell2012gaussification,furrer2018repeaters}, nonlocal ghost-displacement operations for covert information sharing~\cite{zanforlin2023covert}, and a range of quantum state preparation protocols~\cite{pegg1998optical,dakna1999generation,pryde2003creation}. Moreover, the role of measuring ``nothing'' in systems that are continuously monitored has implications more broadly in quantum science~\cite{plenio1998quantum}. 
For example, beyond purely optical systems, zero-photon detection can drive atoms towards decoherence free subspaces in cavity QED~\cite{beige2000quantum,beige2000driving}, dark state preparation in atomic systems~\cite{dalibard1992wave}, developing enhanced quantum sensing techniques~\cite{ilias2022criticality,clark2022exploiting}, speeding up simulations of photon-counting experiments~\cite{wein2024simulating}, and understanding the thermodynamics of nonequllibrium systems~\cite{manzano2022quantum}. Excitingly, zero-photon detection has been recently used to modify optical states~\cite{nunn2021heralding,nunn2022modifying} and the nonclassicality of such states can also be investigated via zero-photon detection~\cite{nunn2023transforming}.

\begin{figure}
    \centering
    \includegraphics[width=\columnwidth]{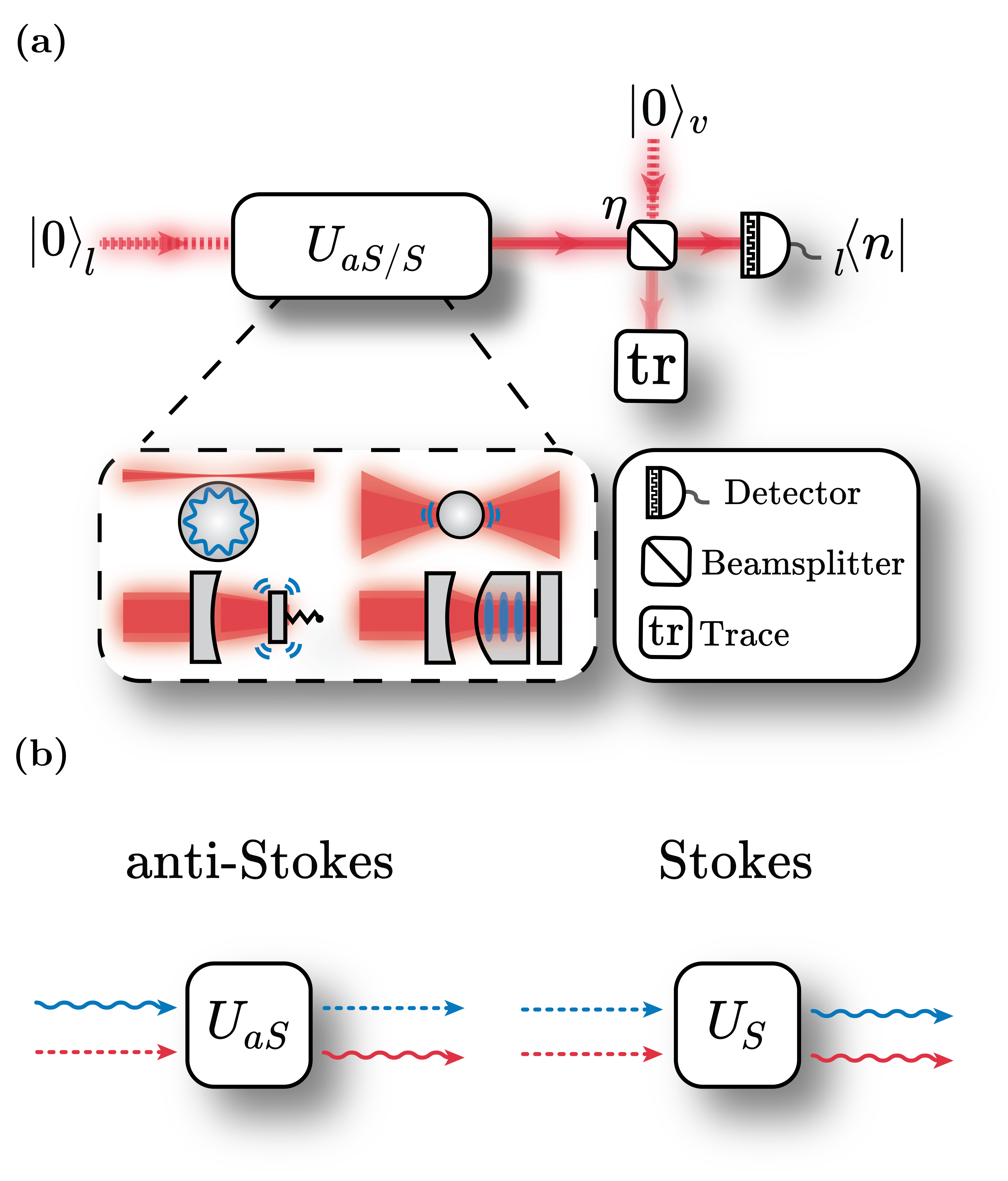}
    \caption{\label{fig:1}(a) Schematic for enhancing or enabling mechanical cooling via zero-photon detection. Here, a vacuum optical mode $\ket{0}_{l}$ interacts with a mechanical mode via the anti-Stokes $(U_{aS})$ and Stokes $(U_{S})$ interactions before impinging on a photon counter. The total optical loss after the cavity and measurement efficiency are modelled using a beamsplitter model for loss with transmission coefficient $\eta$. Here, the initial state of the optical environment is the vacuum state $\ket{0}_{v}$ and one output of the beamsplitter is traced over. These optomechanical interactions can take place in a wide range of architectures including, clockwise from top left of inset, whispering-gallery-mode microresonators, levitated nanoparticles, bulk acoustic resonators, and moving-end-mirror cavities. (b) Depiction of the anti-Stokes and Stokes interactions, where red wavy lines indicate optical photons, blue wavy lines indicate mechanical phonons, and dashed lines indicate vacua. In the anti-Stokes process, an excitation can be coherently swapped from the mechanical to the optical field, while in the Stokes process, pairs of optomechanical excitations are generated.}
    \label{fig:1}
\end{figure}
 
Here, we theoretically explore how zero-photon detection can be used to cool a mechanical oscillator. 
Firstly, we discuss how the mechanical mode can be cooled via zero-photon detection following the anti-Stokes interaction. For any non-zero detection efficiency, we show that this heralded mechanical cooling outperforms standard laser-cooling techniques as was experimentally demonstrated in the companion article to the present work, Ref.~\cite{major2023something}.
Secondly, we investigate the Stokes scattering process, which is typically associated with parametric amplification and heating of the mechanical mode~\cite{marquardt2009optomechanics}. However, we show that zero-photon detection on the Stokes signal can also enable mechanical cooling and we determine the detection efficiency requirements to achieve cooling in this scenario.  
Our theoretical description of mechanical cooling via zero-photon detection is developed for both pulsed measurements and for continuous monitoring, where the latter incorporates the influence of both photon counting and open-system dynamics~\cite{wiseman2009quantum}. As the anti-Stokes and Stokes processes may be realized throughout a wide range of optomechanical systems, including both radiation-pressure-based and Brillouin-scattering-based optomechanics, our scheme for enhanced mechanical cooling is widely applicable. Beyond heralded cooling of a mechanical oscillator, the methods introduced here will impact the development of key areas such as reaching the quantum ground state, thermometry and quantum thermodynamics~\cite{mehboudi2019thermometry}, and mechanical-state tomography~\cite{vanner2015towards}.

\section{Pulsed mechanical cooling via zero-photon detection}
In this section, we study heralded state preparation of a mechanical oscillator via a single zero-photon detection event. The mechanical state is prepared by driving the anti-Stokes or Stokes scattering process with a pulse of light, which is then followed by a photon counting measurement, as illustrated in Fig.~\hyperref[fig:1]{1(a)}. In this section, we focus on mechanical state preparation via short optomechanical interactions, in which the pulse duration $\tau$ is much smaller than the mechanical coherence time, and so the mechanical open-system dynamics may be ignored. An open-quantum systems approach with continuous driving and monitoring is considered in Section~\ref{sec:SME_non_adiabatic}. Important to this work, we consider the effect of optical detection inefficiencies on the ability to cool the mechanical oscillator via zero-photon detection.

The anti-Stokes process is described by the optomechanical beamsplitter Hamiltonian
\begin{equation}
    H_{aS}=\hbar G\left(a\bd+\ad b\right),
\end{equation}
where $a$ ($b$) is the annihilation operator of the optical (mechanical) mode and $G$ is the linearized optomechanical coupling rate. Similarly, the Stokes process is described by the optomechanical two-mode squeezing Hamiltonian
\begin{equation}
    H_{S}=\hbar G\left(\ad\bd+ab\right).
\end{equation}
The anti-Stokes and Stokes scattering processes, graphically depicted in Fig.~\hyperref[fig:1]{1(b)}, can be experimentally realized throughout a wide range of optomechanical systems: from whispering-gallery-mode microresonators to levitation-based optomechanics, and from bulk acoustic resonators to moveable-end-mirror cavities~\cite{meystre2013short}, as indicated in Fig.~\hyperref[fig:1]{1(a)}.  

In the absence of open-system and cavity dynamics, driving the anti-Stokes interaction for a pulse duration $\tau$ realizes the optomechanical unitary $U_{aS}=\rme^{-\rmi H_{aS}\,\tau/\hbar}$ and driving the Stokes interaction achieves $U_{S}=\rme^{-\rmi H_{S}\,\tau/\hbar}$. 
Notably, the optical and mechanical field operators can be characterized by the underlying SU(2) and SU(1,1) Lie algebras for the anti-Stokes and Stokes interactions, respectively~\cite{yurke19862}. These underlying group structures enable the unitaries $U_{aS}$ and $U_{S}$ to be decomposed into other compact and convenient forms~\cite{klimov2009group}, which are utilized in Sections~\ref{sec:no_open_as} and \ref{sec:no_open_s}. Furthermore, in what follows, we will consider the effect of these unitary interactions, followed by photon counting measurements, on an initial thermal mechanical state described by the density operator
\begin{equation}
    \rho_{\bar{n}}=\frac{1}{1+\bar{n}}\sum_{m=0}^{\infty}\left(\frac{\bar{n}}{1+\bar{n}}\right)^m\ket{m}\bra{m}
\end{equation}
with initial thermal occupation $\bar{n}$.

\subsection{Pulsed anti-Stokes interaction}\label{sec:no_open_as} 
Following the anti-Stokes interaction described by $U_{aS}$, a photon counting measurement is performed. The detection of $n$ photons makes a projective measurement of the optical mode onto the $n^{\mathrm{th}}$ Fock state $\ket{n}_{l}$, where the the label $l$ indicates the optical mode. Thus, driving the anti-Stokes process for a pulse duration $\tau$ followed by an $n$-photon detection event can be described by the measurement operator $\Upsilon_{n}^{(aS)}=\prescript{}{l}{\bra{n}}U_{aS}\ket{0}_{l}$. 
Here, we also assume that the initial anti-Stokes mode is the optical vacuum state $\ket{0}_{l}$, which is valid in the linearized regime of both radiation-pressure-based optomechanics and Brillouin-scattering-based optomechanics. Explicitly, the measurement operator $\Upsilon_{n}^{(aS)}$ may be written as
\begin{equation}
    \Upsilon_{n}^{(aS)}=\dfrac{(-R^{*})^n}{\sqrt{n!}}T^{\bd b}b^{n},\label{eq:Upsilon_as_n}
\end{equation}
where $T=\cos(G\tau)$ and $R=-\rmi\sin(G\tau)$, which may be confirmed using the following decomposition for the SU(2) group: $U_{aS}=T^{\bd b}\rme^{-R^{*}\ad b}\rme^{R\bd a}T^{-\ad a}$~\cite{dakna1998photon}. Notably, the operator $b^n$ in Eq.~\eqref{eq:Upsilon_as_n} describes the operation of $n$-fold phonon subtraction. By itself, this operation increases the initial occupation of a mechanical thermal state from $\bar{n}$ to $(n+1)\bar{n}$~\cite{kim2005nonclassicality,bogdanov2017multiphoton,barnett2018statistics}. However, we will now analyze the action of the full measurement operator $\Upsilon_{n}^{(aS)}$ operation including $T^{\bd b}$. Crucially, Eq.~\eqref{eq:Upsilon_as_n} includes terms beyond first order in $G$ and so the action of $T^{\bd b}$ is nontrivial.

After the anti-Stokes interaction and photon-number measurement, the mechanical state is described by the mapping $\rho_{n}^{(aS)}=\Upsilon_{n}^{(aS)}\rho_{\bar{n}}\Upsilon_{n}^{\dagger\,(aS)}/\mathcal{P}_{n}^{(aS)}$, where $\mathcal{P}_{n}^{(aS)}=\tr\left(\Upsilon_{n}^{\dagger\,(aS)}\Upsilon_{n}^{(aS)}\rho_{\bar{n}}\right)$ is the probability of detecting $n$ photons. Using Eq.~\eqref{eq:Upsilon_as_n}, one finds that
\begin{eqnarray}
\rho^{(aS)}_{n}&=&\left(\dfrac{1+\bar{n}\sin^2(G\tau)}{1+\bar{n}}\right)^{n+1}\nonumber\\&&\sum_{m=0}^{\infty}\left(\dfrac{\bar{n}\cos^2(G\tau)}{1+\bar{n}}\right)^{m}\begin{pmatrix}
m+n\\n
\end{pmatrix}\ket{m}\bra{m}\label{eq:rho_as_n}
\end{eqnarray}
and
\begin{equation}
\mathcal{P}_{n}^{(aS)}=\dfrac{\left(\bar{n}\sin^2(G\tau)\right)^n}{\left(1+\bar{n}\sin^2(G\tau)\right)^{n+1}}.
\end{equation}
The occupation of the mechanical state $\rho^{(aS)}_{n}$ can then be calculated via Eq.~\eqref{eq:rho_as_n}, which gives 
\begin{align}
    \tr\left(\bd b \rho^{(aS)}_{n}\right)%
    &=\left(n+1\right)\bar{n}_{0}^{(aS)},\label{eq:rho_as_n_occupation}
\end{align}
where $\bar{n}_{0}^{(aS)}={\bar{n}\cos^2(G\tau)}/\left({1+\bar{n}\sin^2(G\tau)}\right)$. 
For $n=0$, we may therefore conclude that $\bar{n}_{0}^{(aS)}$ in Eq.~\eqref{eq:rho_as_n_occupation} differs from $\bar{n}$ because of the operator $T^{\bd b}$ in Eq.~\eqref{eq:Upsilon_as_n}. Thus, the operator $T^{\bd b}$ acts to reduce the mean phonon number due to the knowledge gained about the mechanical mode via the optomechanical interaction and zero-photon measurement. This Bayesian update from the operator $T^{\bd b}$ is present for any $n$-photon detection event and competes with the $n$-fold phonon subtraction operation $b^{n}$, which increases the mechanical occupation. 

\begin{figure*}
    \centering
    \includegraphics[width=\textwidth]{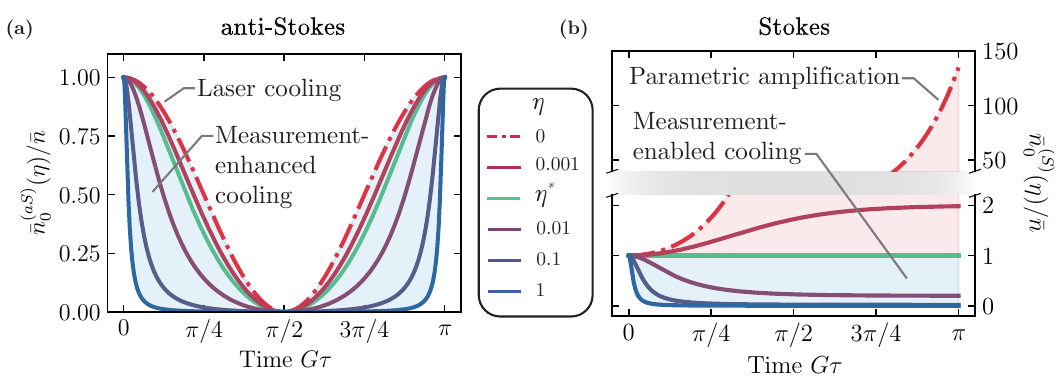}
    \caption{Pulsed zero-photon-heralded cooling and the effect of optical detection inefficiencies. (a) Mechanical cooling via zero-photon detection on the anti-Stokes signal as a function of time for various detection efficiencies $\eta$. Here, $\bar{n}_{0}^{(aS)}$ is the mechanical occupation after zero-photon detection, $\bar{n}$ is the initial mechanical occupation, $G$ is the optomechanical coupling rate, and ${\eta}=0$ corresponds to standard laser cooling. The improved measurement-enhanced cooling with increasing ${\eta}$ is observed and is indicated by the blue shaded region. At times when $G\tau$ is an odd multiple of $\pi/2$, a full state swap has occurred between the vacuum state and the mechanical mode and hence $\bar{n}_{0}^{(aS)}=0$ for all ${\eta}$. Likewise, when $G\tau$ is an even multiple of $\pi/2$, the population is transferred back onto the mechanical oscillator and so $\bar{n}_{0}^{(aS)}=\bar{n}$ for all ${\eta}$.
    (b) Zero-photon detection on the Stokes signal as a function of time for various detection efficiencies. Here, $\bar{n}_{0}^{(S)}$ is the mechanical occupation after zero-photon detection and ${\eta}=0$ corresponds to mechanical heating via optomechanical parametric amplification. Measurement-enabled mechanical cooling is achieved when ${\eta}$ is above the value of ${\eta}^*=1/\left(1+\bar{n}\right)$ and the mechanical occupation tends to $(1-{\eta})/{\eta}$ as $G\tau$ increases. The measurement-enabled cooling region is shaded light blue and the region where heating occurs is shaded light red. In plots (a) and (b) the occupation of the initial thermal state prior to the optomechanical interaction is $\bar{n}=500$, which gives a threshold efficiency of ${\eta}^*\approx 0.002$ for the Stokes interaction.}
    \label{fig:2}
\end{figure*}

\subsubsection{Cooling via zero-photon detection and the anti-Stokes interaction}
We will now show how averaging over all photon-counting measurement outcomes leads to laser cooling and compare this to the enhanced cooling effect achieved via zero-photon detection. The detection of zero photons after a pulsed anti-Stokes interaction of duration $\tau$ corresponds to setting $n=0$ in Eq.~\eqref{eq:rho_as_n}, which gives the mechanical state 
\begin{equation}
    \rho_{0}^{(aS)}=\frac{1}{1+\bar{n}_{0}^{(aS)}}\sum_{m=0}^{\infty}\left(\frac{\bar{n}_{0}^{(aS)}}{1+\bar{n}_{0}^{(aS)}}\right)^m\ket{m}\bra{m}.\label{eq:rho_as_0}
\end{equation}
$\rho_{0}^{(aS)}$ takes the form of a thermal state, and $\bar{n}_{0}^{(aS)}$ is the occupation of the thermal state following the interaction and measurement.  

This heralded cooling of the mechanical mode via zero-photon detection can be compared to deterministic laser cooling, which corresponds to no optical measurement being made after the anti-Stokes interaction described by $\rho_{LC}=\sum_{n}\mathcal{P}_{n}^{(aS)}\rho_{n}^{(aS)}$. Eq.~\eqref{eq:rho_as_n} and the associated probabilities $\mathcal{P}_{n}^{(aS)}$ can be used to show that  $\rho_{LC}$ is also a thermal state but with thermal occupation now given by $\bar{n}_{LC}=\bar{n}\cos^2(G\tau)$. For this model, for $G\tau=(2k+1)\pi/2$ with $k\in\mathbb{N}$ one has $\bar{n}^{(aS)}_{0}=\bar{n}_{LC}=0$, which corresponds to a full state swap between the optics (initially in vacuum) and the mechanical thermal state. Whereas, at $G\tau=k\pi$ with $k\in\mathbb{N}$ one has $\bar{n}^{(aS)}_{0}=\bar{n}_{LC}=\bar{n}$ and so, for $k>0$, all the population has been transferred back onto the mechanics. However, for all other values of $G\tau$, one has $\bar{n}^{(aS)}_{0}<\bar{n}_{LC}$, and thus, conditional cooling via zero-photon detection exceeds the limits of standard laser cooling as shown in Fig.~\ref{fig:2}(a). 

The inclusion of optical detection inefficiencies allows one to observe the transition between these two cooling strategies. Namely, when the detection efficiency is zero---which effectively traces out the output optical mode---we have the laser cooling case. To describe such detection losses we employ a beamsplitter model for loss~\cite{leonhardt1997measuring}. In this model, a beamsplitter $B$ with intensity transmission coefficient ${\eta}$ couples the output optical mode to an environmental vacuum mode, labelled by $v$, just before photon counting. Following this beamsplitter interaction, a trace operation is performed over the environmental mode to describe the attenuation of the anti-Stokes signal as is shown in Fig.~\ref{fig:1}(a). Thus, we may calculate the mechanical state in the presence of detection inefficiencies according to
\begin{eqnarray}    
 \rho_{0}^{(aS)}\left({\eta}\right) =\dfrac{\sum_{m=0}^{\infty}\Upsilon_{0,m}^{(aS)}\rho_{\bar{n}}\Upsilon_{0,m}^{\dagger\,(aS)}}{\sum_{m=0}^{\infty}\tr\left(\Upsilon_{0,m}^{\dagger\,(aS)}\Upsilon_{0,m}^{(aS)}\rho_{\bar{n}}\right)},\label{eq:rho_as_0_loss}
\end{eqnarray}
where the measurement operator $\Upsilon_{0,m}^{(aS)}$ represents the operation on the mechanics resulting from zero photons being detected and $m$ photons being lost to the vacuum environment. This measurement operator is given by $\Upsilon_{0,m}^{(aS)}=\prescript{}{v}{\bra{m}}\prescript{}{l}{\bra{0}}BU_{aS}\ket{0}_{l}\ket{0}_{v}$ and, by decomposing the beamsplitter unitary $B$ using the same decomposition formula for the SU(2) group as above, one finds that the state $\rho_{0}^{(aS)}$ in Eq.~\eqref{eq:rho_as_0_loss} is a mechanical thermal state with an occupation now given by
\begin{eqnarray}
    \bar{n}_{0}^{(aS)}\left({\eta}\right) =\dfrac{\bar{n}\cos^2(G\tau)}{{1+{\eta}\bar{n}\sin^2(G\tau)}}.\label{eq:n0_as_loss}
\end{eqnarray}
In Fig.~\ref{fig:2}(a), we plot Eq.~\eqref{eq:n0_as_loss} as function of pulsed interaction strength $G\tau$ for various values of ${\eta}$. Hence, we confirm that at ${\eta}=0$, or $G\tau=(2k+1)\pi/2,~k\pi$ with $k\in\mathbb{N}$,  $\bar{n}_{0}^{(aS)}=\bar{n}_{LC}$. Also, for all other values of $G\tau$ and $0<{\eta}\leq1$, one has that $\bar{n}_{0}^{(aS)}\left({\eta}\right) <\bar{n}_{LC}<\bar{n}$. Thus, the heralded cooling effect from a zero-click detection event is stronger than standard laser cooling even in the presence of detection inefficiencies. The probability for a zero-photon detection event is given by $\mathcal{P}_{0}^{(aS)}\left({\eta}\right) =\sum_{m=0}^{\infty}\tr\left(\Upsilon_{0,m}^{\dagger\,(aS)}\Upsilon_{0,m}^{(aS)}\rho_{\bar{n}}\right)=1/\left(1+{\eta}\bar{n}\sin^2(G\tau)\right)$ [cf. Eq.~\eqref{eq:rho_as_0_loss}].

\subsection{Pulsed Stokes interaction}\label{sec:no_open_s}
The unconditional evolution of the mechanical mode during the Stokes interaction corresponds to heating via optomechanical parametric amplification. By performing photon-counting measurements on the output Stokes signal, phonon-addition operations can be heralded~\cite{vanner2013quantum}, which also increase the mean mechanical occupation of an initial thermal state. Here, we discuss how utilizing zero-photon detection events helps to suppress the heating effects of the Stokes interaction and even provides measurement-enabled mechanical cooling.

Driving the Stokes interaction $U_{S}$ for a time $\tau$ and then detecting $n$ photons is described by the measurement operator $\Upsilon_{n}^{(S)}=\prescript{}{l}{\bra{n}}U_{S}\ket{0}_{l}$, which is explicitly given by
\begin{equation}
    \Upsilon_{n}^{(S)}= \dfrac{\mathcal{R}^{n}}{\sqrt{n!}}\mathcal{T}^{-\left(\bd b + 1\right)}\left(\bd\right)^{n},\label{eq:Upsilon_s_n}
\end{equation}
where $\mathcal{T} = \cosh(G\tau)$, $\mathcal{R} = -\rmi\sinh(G\tau)$, and the decomposition formula for the SU(1,1) group $U_{S}= \rme^{\mathcal{R}\ad \bd/\mathcal{T}} \mathcal{T}^{-(\ad a + \bd b + 1)} \rme^{\mathcal{R}ab/\mathcal{T}}$ was used~\cite{schumaker1985new}. The mechanical state following the Stokes interaction and $n$-photon detection event is $\rho_{n}^{(S)}=\Upsilon_{n}^{(S)}\rho_{\bar{n}}\Upsilon_{n}^{\dagger\,(S)}/\mathcal{P}_{n}^{(S)}$, where $\mathcal{P}_{n}^{(S)}=\tr\left(\Upsilon_{n}^{\dagger\,(S)}\Upsilon_{n}^{(S)}\rho_{\bar{n}}\right)$ is the probability to detect $n$ photons. Using Eq.~\eqref{eq:Upsilon_s_n} then allows one to arrive at
\begin{eqnarray}
\rho^{(S)}_{n}&=&\left(\frac{\left(\cosh^2(G\tau)+\bar{n}\sinh^2(G\tau)\right)^{n+1}}{\bar{n}^n\left(1+\bar{n}\right)\cosh^2(G\tau)}\right)\nonumber\\&&\sum_{m=n}^{\infty}\left(\frac{\bar{n}}{\left(1+\bar{n}\right)\cosh^2(G\tau)}\right)^m\begin{pmatrix}
m\\n
\end{pmatrix}\ket{m}\bra{m}\label{eq:rho_s_n}
\end{eqnarray}
and  
\begin{equation}
    \mathcal{P}_{n}^{(S)}= \dfrac{\left(\left(1+\bar{n}\right)\sinh^2(G\tau)\right)^n}{\left(1+\left(\bar{n}+1\right)\sinh^2(G\tau)\right)^{n+1}}.
\end{equation}
The mean occupation of $\rho^{(S)}_{n}$ can then be calculated via Eq.~\eqref{eq:rho_s_n}, which gives 
\begin{align}
    \tr\left(\bd b \rho^{(S)}_{n}\right)
    &=\left(n+1\right)\bar{n}_{0}^{(S)}+n,\label{eq:rho_s_n_occupation}
\end{align}
where $\bar{n}_{0}^{(S)}=\bar{n}/\left(1+(\bar{n}+1)\sinh^2(G\tau)\right)$. The non-Gaussian operation of $n$-fold phonon addition increases the initial thermal occupation from $\bar{n}$ to $\left(n+1\right)\bar{n}+n$~\cite{agarwal1991nonclassical,barnett2018statistics}, which is equivalent to Eq.~\eqref{eq:rho_s_n_occupation} when $\bar{n}_{0}^{(S)}$ is replaced with $\bar{n}$. Thus, the mathematical difference between $n$-phonon addition and $n$-click detection on the Stokes signal originates from the operator $\mathcal{T}^{-\bd b}$ in Eq.~\eqref{eq:Upsilon_s_n}.

\subsubsection{Cooling via zero-photon detection and the Stokes interaction}
Detecting zero photons after a pulsed Stokes interaction produces the mechanical state in Eq.~\eqref{eq:rho_s_n} with $n=0$, which is 
\begin{equation}
    \rho_{0}^{(S)}=\frac{1}{1+\bar{n}_{0}^{(S)}}\sum_{m=0}^{\infty}\left(\frac{\bar{n}_{0}^{(S)}}{1+\bar{n}_{0}^{(S)}}\right)^m\ket{m}\bra{m}.\label{eq:rho_s_0}
\end{equation}
Once again, $\rho_{0}^{(S)}$ takes the form of a thermal state with thermal occupation $\bar{n}_{0}^{(S)}$. In the absence of optical loss, for $G\tau>0$ the thermal occupation $\bar{n}_{0}^{(S)}$ is always less than the initial thermal occupation $\bar{n}$. Thus, even when the mechanical mode is driven by the Stokes interaction, which unconditionally heats the mechanics via the optomechanical two-mode squeezing interaction, zero-photon detection enables conditional mechanical cooling. 

The unconditional heating of the mechanical mode via the Stokes interaction is described by the density operator $\rho_{TMS}=\sum_{n}\mathcal{P}_{n}^{(S)}\rho_{n}^{(S)}$. Similar to the anti-Stokes case, Eq.~\eqref{eq:rho_s_n}, and the expression for $\mathcal{P}_{n}^{(S)}$, can be used to show that $\rho_{TMS}$ is also a thermal state with a thermal occupation $\bar{n}_{TMS}=\bar{n}+(\bar{n}+1)\sinh^2(G\tau)$. As in Section~\ref{sec:no_open_as}, the effect of optical detection inefficiencies on mechanical cooling via zero-photon detection can also be calculated via a beamsplitter model for loss. In this way, the mechanical state after a zero-photon detection event in the presence of detection inefficiencies is 
\begin{eqnarray}    
 \rho_{0}^{(S)}\left({\eta}\right) 
 =\dfrac{\sum_{m=0}^{\infty}\Upsilon_{0,m}^{(S)}\rho_{\bar{n}}\Upsilon_{0,m}^{\dagger\,(S)}}{\sum_{m=0}^{\infty}\tr\left(\Upsilon_{0,m}^{\dagger\,(S)}\Upsilon_{0,m}^{(S)}\rho_{\bar{n}}\right)},\label{eq:rho_s_0_loss}
\end{eqnarray}
where the measurement operator is now given by $\Upsilon_{0,m}^{(S)}={}_{v}\bra{m}{}_{l}\bra{0}BU_{S}\ket{0}_{l}\ket{0}_{v}$. By following the same steps outlined above for the anti-Stokes case, one can show that $\rho_{0}^{(S)}$ in Eq.~\eqref{eq:rho_as_0_loss} is a mechanical thermal state with occupation
\begin{eqnarray}
    \bar{n}_{0}^{(S)}\left({\eta}\right)=\dfrac{\bar{n}+\left(1+\bar{n}\right)\left(1-{\eta}\right)\sinh^2(G\tau)}{1+{\eta}\left(1+\bar{n}\right)\sinh^2(G\tau)}\label{eq:n0_s_loss}
\end{eqnarray}
and the associated probability for a zero-photon detection event is $\mathcal{P}_{0}^{(S)}\left({\eta}\right) =\sum_{m=0}^{\infty}\tr\left(\Upsilon_{0,m}^{\dagger\,(S)}\Upsilon_{0,m}^{(S)}\rho_{\bar{n}}\right)=1/\left(1+{\eta}\left(\bar{n}+1\right)\sinh^2(G\tau)\right)$. Hence, we confirm that for ${\eta}=0$ one recovers the result $\bar{n}_{0}^{(S)}=\bar{n}_{TMS}$. In Fig.~\ref{fig:2}(b), we plot Eq.~\eqref{eq:n0_s_loss} as a function of time for various values of ${\eta}$. Interestingly, we find that there exists a threshold value for ${\eta}$ given by ${\eta}^*=1/\left(\bar{n}+1\right)$ where the effects of parametric amplification and cooling via zero-photon detection balance. When ${\eta}$ is above this value, the cooling effect via zero-photon detection outweighs the deterministic heating contribution and the mechanical state is cooled conditionally via zero-photon detection. Notably, for mechanical systems with $\bar{n}\gg1$, ${\eta}^*\approx0$ and so conditional mechanical cooling via zero-photon detection on the Stokes signal is readily achievable for typical room temperature mechanical oscillators. However, for mechanical systems approaching the ground state, where $\bar{n}$ is close to zero, ${\eta}^*$ tends to unity. Hence, conditional cooling via the anti-Stokes signal becomes more favourable than via the Stokes signal for near-ground-state systems when the optical detection efficiency is a limiting factor. Finally we note that for $\eta>0$, as $G\tau\rightarrow\infty$, $\bar{n}_{0}^{(S)}\left({\eta}\right)\rightarrow\left(1-{\eta}\right)/{\eta}$ which is finite despite $\bar{n}_{TMS}\rightarrow\infty$.

\section{Continuous monitoring and open-system dynamics}\label{sec:SME_non_adiabatic}
\begin{figure}
    \centering
        \includegraphics[width=\columnwidth]{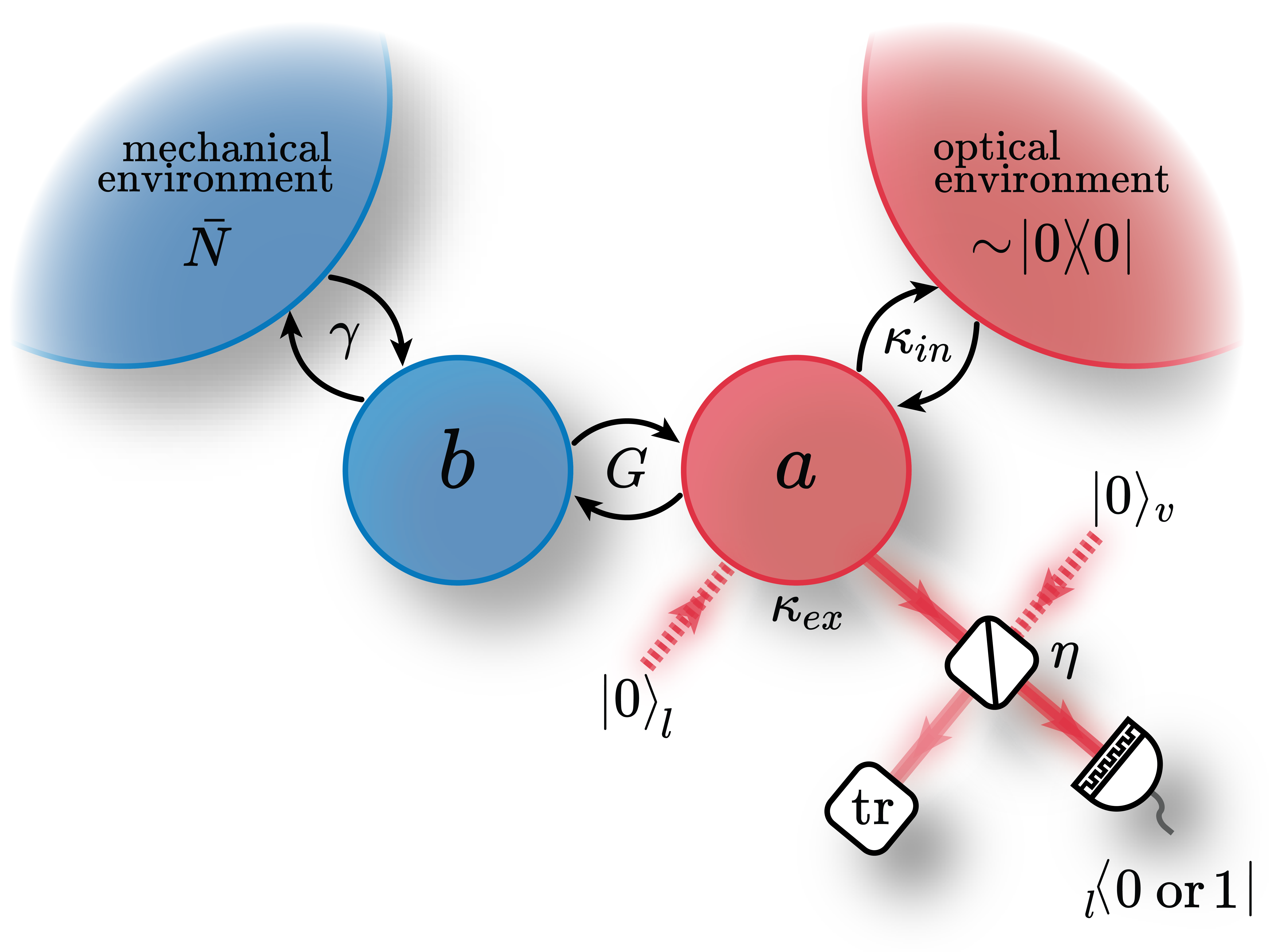}%
    \caption{Open-system and cavity optomechanical dynamics in a continuously driven system monitored by imperfect photon counting measurements. Here, the optical $(a)$ and mechanical $(b)$ modes interact in a cavity via the anti-Stokes or Stokes interaction with optomechanical coupling strength $G$. The cavity couples to the optical vacuum input $\ket{0}_{l}$ and the output field at rate $\kappa_{ex}$. A photon counter continuously measures the cavity output, which produces a random string of outcomes comprised of zero- and single-photon detection events. The mechanical mode interacts at a rate $\gamma$ with a thermal environment with occupation $\bar{N}$ and the optical cavity mode additionally decays into a vacuum environment at a rate $\kappa_{in}$. Meanwhile, detection inefficiencies are modelled via a beamsplitter model for loss prior to photon counting. 
    }
    \label{fig:3}
\end{figure}

We now extend our description to a continuously monitored optomechanical system, in which both mechanical open-system and cavity dynamics are included as depicted in Fig.~\ref{fig:3}. Here, an optical cavity mode is driven by a continuous pump laser to drive the anti-Stokes or Stokes interaction while photon-counting measurements are made continuously at the cavity output on the anti-Stokes or Stokes signal, respectively. This situation is described by the following stochastic master equation (SME) for the joint state $\rho$ of the cavity and mechanical mode:
\begin{align}
    d{\rho}=&-\frac{i}{\hbar}[H,\rho]{dt}+\mathcal{G}[a]\rho{dN}-{\eta}\kappa_{ex}\mathcal{H}\left[ a^{\dagger} a\right]\rho{dt}\nonumber\\
    &+2\kappa_{ex}(1-{\eta})\mathcal{D}\left[a\right]\rho{dt}\nonumber+2\kappa_{in}\mathcal{D}\left[a\right]\rho{dt}\nonumber\\
    &+2\gamma(\bar{N}+1)\mathcal{D}[b]\rho{dt}+2\gamma\bar{N}\mathcal{D}[b^{\dagger}]\rho{dt},\label{eq:SME}
\end{align}
which assumes the photon-counting duration is smaller than all other relevant timescales~\cite{caves1987quantum}. To describe the anti-Stokes or Stokes scattering processes, one chooses $H=H_{aS}$ or $H=H_{S}$ in Eq.~\eqref{eq:SME}, respectively, and the superoperators are given by $\mathcal{G}[O]\rho=O\rho{O}^{\dagger}/\expval{O^{\dagger}O}-\rho$, $\mathcal{H}[O]\rho=O\rho+\rho{O^{\dag}}-\expval{O+O^{\dagger}}\rho$, and $\mathcal{D}[O]\rho=O\rho{O^{\dag}}-\frac{1}{2}\{O^{\dag}O,\rho\}$. The parameters characterizing the open-system and cavity dynamics in Eq.~\eqref{eq:SME} are shown in Fig.~\ref{fig:3}, which are the external (intrinsic) amplitude decay rate of the cavity mode $\kappa_{ex}~(\kappa_{in})$,  the amplitude decay rate of the mechanical mode $\gamma$, the detection efficiency of photon counting ${\eta}$, and the occupation of the mechanical thermal environment $\bar{N}$. Importantly, photon counting is a stochastic process and so the stochastic increment may take the value ${dN}=0$ or ${dN}=1$ for a zero-photon or single-photon detection event, respectively, and so ${dN}^2={dN}$. Zero-photon and single-photon detection occurs with probabilities $\mathcal{P}_{0}=1-2{\eta}\kappa_{ex}\expval{\ad a}\rmd{t}$ and $\mathcal{P}_{1}=2{\eta}\kappa_{ex}\expval{\ad a}\rmd{t}$ during each $\rmd{t}$. To find the total probability of a particular measurement record, one may then multiply the corresponding probabilities at every instant in time. We note that in the construction of Eq.~\eqref{eq:SME} it is assumed that the measurement of two or more photons during each $dt$ occurs with probability of order $dt^2$ or higher, which therefore do not contribute to the SME. 

The equation describing the evolution of an expectation value of an arbitrary operator $O$ in $a$, $b$ may be calculated via Eq.~\eqref{eq:SME} and $d\expval{O}=\tr\left(Od\rho\right)$~\cite{jacobs2006straightforward,*jacobs2014quantum}, which gives
\begin{gather}
    d{\expval{O}}=\frac{i}{\hbar}\expval{[H,O]}\rmd{t}+\left[\dfrac{\expval{\ad O a}}{\expval{\ad a}}-\expval{O}\right]{dN}\nonumber\\
    -{\eta}\kappa_{ex}\left[\expval{O\ad a}+\expval{\ad a O}-2\expval{O}\expval{\ad a}\right]{dt}\nonumber\\
    +\left[2{\kappa}-2{\eta}\kappa_{ex}\right]\left[\expval{\ad O a}-\frac{1}{2}\expval{O\ad a}-\frac{1}{2}\expval{\ad a O}\right]{dt}\nonumber\\
    +2\gamma(\bar{N}+1)\left(\expval{\bd O b}-\frac{1}{2}\expval{O\bd b}-\frac{1}{2}\expval{\bd b O}\right) dt\nonumber\\
    +2\gamma\bar{N}\left(\expval{b O \bd}-\frac{1}{2}\expval{Ob \bd}-\frac{1}{2}\expval{ b \bd O}\right)dt,\label{eq:ajoint_SME_full}
\end{gather}
where ${\kappa}=\kappa_{ex}+\kappa_{in}$ is the total amplitude cavity decay rate.

\subsection{Continuous anti-Stokes interaction}\label{sec:open_as}

Here, we choose $H=H_{aS}$ in Eq.~\eqref{eq:ajoint_SME_full} to select the anti-Stokes scattering process, which enables an open-quantum-systems description of: \hyperlink{link:as_LC}{(i)}~mechanical laser cooling, \hyperlink{link:as_ZC}{(ii)}~enhanced laser cooling via zero-photon detection, and \hyperlink{link:as_1C}{(iii)}~single-photon detection events.
\hypertarget{link:as_LC}{}\subsubsection{Laser cooling}\label{sec:open_as_lc}

Laser cooling is realized in the absence of photon counting, which corresponds to setting ${\eta}=0$ and $\rmd{N}=0$ for all time in the equations of motion. By using the canonical commutation relations, $[a,\ad]=1$ and $[b,\bd]=1$, one may use Eq.~\eqref{eq:ajoint_SME_full} to derive the closed set of coupled differential equations describing laser cooling
\begin{align}
    &\rmi\dfrac{d}{\rmd{t}}\expval{\ad b-a\bd}=2G\left[\expval{\ad a}-\expval{\bd b}\right]\nonumber\\
    &\hspace{70pt}-\rmi\left({\kappa}+\gamma\right)\expval{\ad b-a\bd},\label{eq:LC_as_CDE_1}\\
    &\dfrac{d}{\rmd{t}}\expval{\ad a}=-\rmi G \expval{\ad b-a\bd}-2{\kappa}\expval{\ad a},\label{eq:LC_as_CDE_2}\\
    &\dfrac{d}{\rmd{t}}\expval{\bd b}=\rmi G \expval{\ad b-a\bd}-2\gamma\expval{\bd b}+2\gamma\bar{N}.\label{eq:LC_as_CDE_3}
\end{align}
To derive Eqs~\eqref{eq:LC_as_CDE_1}, \eqref{eq:LC_as_CDE_2}, and \eqref{eq:LC_as_CDE_3}, no assumptions need to be made about the initial optical-mechanical state and these coupled differential equations may be written as the first-order matrix differential equation
\begin{eqnarray}
    \dot{\VV}_{aS}&=&A_{aS}\VV_{aS}+\mathbf{n}_{aS},\label{eq:matrix_eqn_1}
\end{eqnarray}
where $\VV_{aS}=\left(\rmi\expval{\ad b-a\bd},~\expval{\ad a },~\expval{\bd b}\right)^{\mathrm{T}}$, $\mathbf{n}_{aS}=\left(0,~0,~2\gamma\bar{N}\right)^{\mathrm{T}}$, and
\begin{eqnarray}
    A_{aS}&=&\begin{pmatrix}-\left({\kappa}+\gamma\right) & 2G & -2G\\
    -G & -2{\kappa} & 0\\
    G & 0 & -2\gamma \end{pmatrix}.\label{eq:Aasmatrix}
\end{eqnarray}
The steady-state solution of Eq.~\eqref{eq:matrix_eqn_1} is
\begin{eqnarray}
\VV_{aS}(t\rightarrow\infty)&=&-A_{aS}^{-1}\mathbf{n}_{aS}\nonumber\\
&=&\begin{pmatrix}-\dfrac{2G\bar{N}\gamma{\kappa}}{(\gamma+{\kappa})(G^2+\gamma{\kappa})}\\  \\ \dfrac{\bar{N}G^2\gamma}{(\gamma+{\kappa})(G^2+\gamma{\kappa})} \\ \\ \dfrac{\bar{N}\gamma[G^2+{\kappa}(\gamma+{\kappa})]}{(\gamma+{\kappa})(G^2+\gamma{\kappa})}\end{pmatrix}.\label{eq:lasercooled_as_long_time}
\end{eqnarray}
Note that in the weak coupling regime, when $G$ is much smaller than ${\kappa}$ and $\gamma$, the last entry of Eq.~\eqref{eq:lasercooled_as_long_time}, which describes the steady state mechanical occupation, is well approximated as $\bar{N}/\left(1+C\right)$. Here, $C=G^2/\left({\kappa}\gamma\right)$ is the optomechanical cooperativity. Finally, the solution of Eq.~\eqref{eq:matrix_eqn_1} with time for the initial condition $\VV_{aS}(0)$ is
\begin{eqnarray}
    \VV_{aS}(t)=-A_{aS}^{-1}\mathbf{n}_{aS}+\rme^{A_{aS} t}\left[\VV_{aS}(0)+A_{aS}^{-1}\mathbf{n}_{aS}\right].\label{eq:laser_cooling_solution_1}
\end{eqnarray}
Note that the real part of the eigenvalues of $A_{aS}$ are negative, which ensures Eq.~\eqref{eq:laser_cooling_solution_1} is stable.

\hypertarget{link:as_ZC}{}\subsubsection{Anti-Stokes interaction and continuous-time zero-photon detection}\label{sec:open_as_zc}

Observing a string of zero-photon detection events is described by setting $\rmd N=0$ and $0<{\eta}\leq1$ in Eq.~\eqref{eq:ajoint_SME_full}. In what follows, we will assume an initial Gaussian optomechanical state~\cite{serafini2017quantum} with zero mean amplitudes $\expval{a}=\expval{\ad}=\expval{b}=\expval{\bd}=0$. Relevant examples of optical-mechanical states satisfying this assumption are: an initial product of optical vacuum and a thermal mechanical state $\ket{0}_{l}\prescript{}{l}{\bra{0}}\otimes\rho_{\bar{n}}$ and the laser-cooled optical-mechanical state in Eq.~\eqref{eq:lasercooled_as_long_time}. This zero-mean Gaussian assumption enables the Isserlis-Wick theorem~\cite{isserlis1918formula,barnett1997methods} to be employed, which states that for a Gaussian state $\expval{ABCD}=\expval{AB}\expval{CD}+\expval{AC}\expval{BD}+\expval{AD}\expval{BC}$ where the expectation values of $A$, $B$, $C$, and $D$ are all zero. The Isserlis-Wick theorem may then be applied to expectation values comprised of the optical and mechanical field operators
${a}$, ${\ad}$, ${b}$, and ${\bd}$ to arrive at the following set of differential equations for the second-order moments:

\begin{widetext}
\begin{align}
    &\rmi\dfrac{d}{\rmd{t}}\expval{\ad b}=G\left[\expval{\ad a}-\expval{\bd b}\right]-\rmi\left({\kappa}+\gamma\right)\expval{\ad b}-2\rmi{\eta}\kappa_{ex}\expval{\ad a}\expval{\ad b}-2\rmi{\eta}\kappa_{ex}\expval{a^{\dagger\,2}}\expval{a b},\label{eq:zc_as_cde_1}\\
    &\rmi\dfrac{d}{\rmd{t}}\expval{a\bd}=-G\left[\expval{\ad a}-\expval{\bd b}\right]-\rmi\left({\kappa}+\gamma\right)\expval{a\bd}-2\rmi{\eta}\kappa_{ex}\expval{\ad a}\expval{a \bd}-2\rmi{\eta}\kappa_{ex}\expval{a^2}\expval{\ad \bd},\label{eq:zc_as_cde_2}\\
    &\dfrac{d}{\rmd{t}}\expval{\ad a}=-\rmi G \expval{\ad b-a\bd}-2{\kappa}\expval{\ad a}-2{\eta}\kappa_{ex}\expval{\ad a}^2-2{\eta}\kappa_{ex}\expval{a^2}\expval{a^{\dagger\,2}},\label{eq:zc_as_cde_3}\\
    &\dfrac{d}{\rmd{t}}\expval{\bd b}=\rmi G \expval{\ad b-a\bd}-2\gamma\expval{\bd b}+2\gamma\bar{N}-2{\eta}\kappa_{ex}\expval{a\bd}\expval{\ad b}-2{\eta}\kappa_{ex}\expval{\ad\bd}\expval{a b},\label{eq:zc_as_cde_4}
\end{align}
\end{widetext}
which are nonlinear owing to the act of measurement via continuous zero-photon detection. Importantly, by assuming an initial state $\ket{0}_{l}\prescript{}{l}{\bra{0}}\otimes\rho_{\bar{n}}$, the terms $\expval{a^2}$, $\expval{a^{\dagger\, 2}}$, $\expval{ab}$, and $\expval{\ad\bd}$ in Eqs~\eqref{eq:zc_as_cde_1} to \eqref{eq:zc_as_cde_4} are zero for all times. More specifically, as these moments are zero for the initial state $\ket{0}_{l}\prescript{}{l}{\bra{0}}\otimes\rho_{\bar{n}}$ and, as shown in Appendix~\ref{appendix:zeroclick_as_1}, the moments $\expval{a^2}$, $\expval{a^{\dagger\, 2}}$, $\expval{ab}$, and $\expval{\ad\bd}$ form a closed set of coupled differential equations, which guarantees $\expval{a^2}=\expval{a^{\dagger\, 2}}=\expval{ab}=\expval{\ad\bd}=0$ for all times. With these terms equalling zero, the equations for $\expval{\ad b}$ and $-\expval{a \bd}$, Eqs~\eqref{eq:zc_as_cde_1} and \eqref{eq:zc_as_cde_2}, are now identical. Thus, if at some initial time $\expval{\ad b}=-\expval{a \bd}$, such as for the initial state $\ket{0}_{l}\prescript{}{l}{\bra{0}}\otimes\rho_{\bar{n}}$, then $\expval{\ad b}=-\expval{a \bd}$ for all times. This observation also allows us to simplify the first element of the vector $\VV_{aS}$ as $\rmi\expval{\ad b-a\bd}=2\rmi\expval{\ad b}$.

With these considerations in mind, the matrix differential equation describing cooling via continuous zero-photon detection is 
\begin{eqnarray}
    \dot{\VV}_{aS}&=&A_{aS}\VV_{aS}+\mathbf{n}_{aS}-{\eta}\kappa_{ex}\mathbf{Z}_{aS},\label{eq:matrix_DE_zero}
\end{eqnarray}
where $\mathbf{Z}_{aS}=\left(4\rmi\expval{\ad b}\expval{\ad a},~2\expval{\ad a}^2,~-2\expval{\ad b}^2\right)^\mathrm{T}$, $A_{aS}$ and $\mathbf{n}_{aS}$ are defined in and above Eq.~\eqref{eq:Aasmatrix}, and now $ \VV_{aS}=\left(2\rmi\expval{\ad b},~\expval{\ad a },~\expval{\bd b}\right)^{\mathrm{T}}$. To obtain the second-order moments as a function of time, we solve Eq.~\eqref{eq:matrix_DE_zero} numerically.

\begin{figure*}
    \centering
    \includegraphics[width=0.8\textwidth]{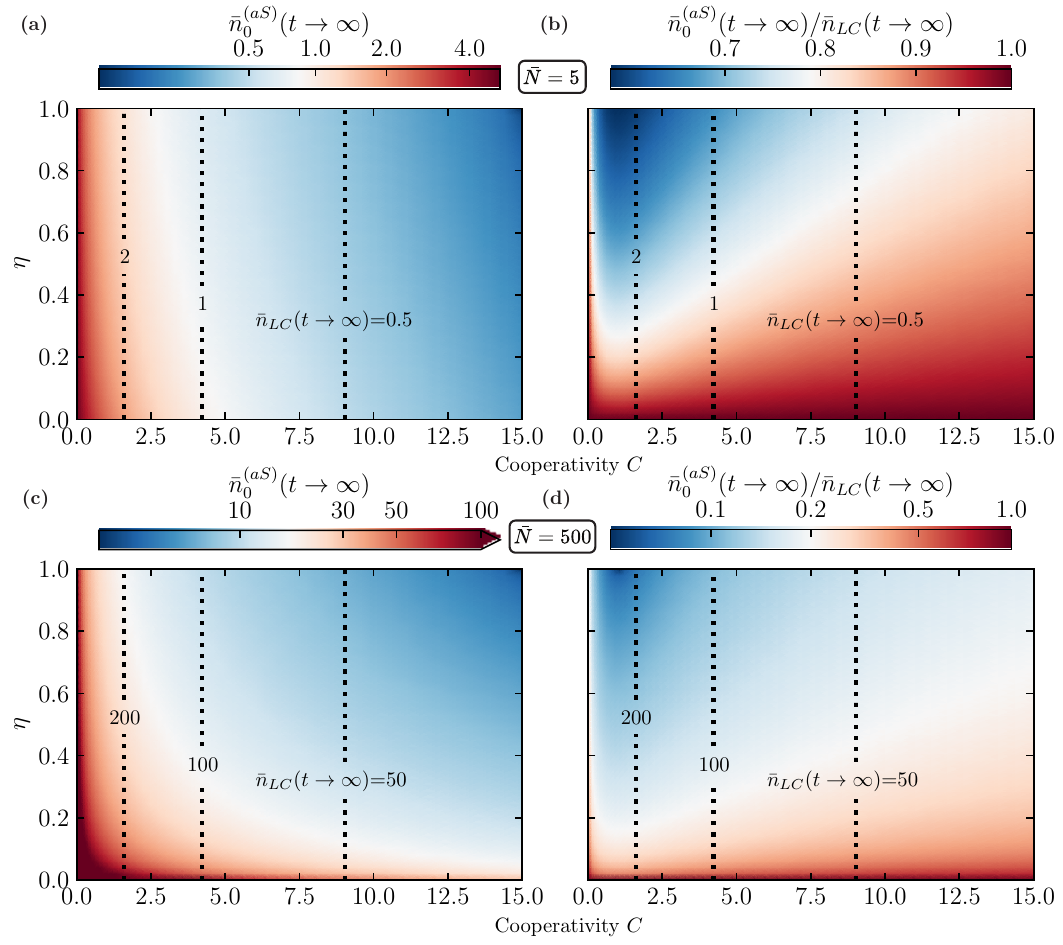}
    \caption{Mechanical cooling via the anti-Stokes interaction and continuous zero-photon detection. (a) The steady-state mechanical occupation after continual zero-photon detection $\bar{n}_{0}^{(aS)}(t\rightarrow\infty)$ as a function of detection efficiency ${\eta}$ and optomechanical cooperativity $C$ for $\bar{N}=5$.
    (b) The ratio between the steady-state mechanical occupations after continual zero-photon detection and unconditioned laser cooling $\bar{n}_{0}^{(aS)}(t\rightarrow\infty)/\bar{n}_{LC}(t\rightarrow\infty)$ for $\bar{N}=5$. Plots (c) and (d) illustrate the same quantities as (a) and (b) respectively but for $\bar{N}=500$. To emphasize the effect of cooling via zero-photon detection, all occupations above 100 are shown as a single colour (dark red). The vertical dashed lines indicate laser-cooled mechanical occupations $\bar{n}_{LC}(t\rightarrow\infty)$. The values of the other parameters used in the simulation are ${\kappa}=\kappa_{ex}=40$, and $\gamma=1$. Here, all rates are normalized by $\gamma$.}
    \label{fig:4}
\end{figure*}

In Fig.~\ref{fig:4}(a) we plot the steady-state mechanical occupation achieved via continuous zero-photon detection $\bar{n}_{0}^{(aS)}(t\rightarrow\infty)$ as a function of detection efficiency $\eta$ and optomechanical cooperativity $C=G^2/\left({\kappa}\gamma\right)$ for $\bar{N}=5$. And in Fig.~\ref{fig:4}(b), we plot the ratio of the continual zero-photon-detection-cooled mechanical occupation $\bar{n}_{0}^{(aS)}(t\rightarrow\infty)$ to the steady-state mechanical occupation achieved via laser cooling $\bar{n}_{LC}(t\rightarrow\infty)$. Fig.~\ref{fig:4} therefore illustrates how zero-photon detection can be used to enhance the level of cooling beyond what is realizable via laser cooling. In particular, the region between $C=0$ and $C=5$ demonstrates that continuous zero-photon detection offers the greatest relative enhancement over laser cooling for the parameter set investigated. 
Moreover, inspecting Fig.~\ref{fig:4}(b) about the vertical line labelled $\bar{n}_{LC}(t\rightarrow\infty)=1$, shows that conditional cooling via zero-photon detection can be utilized to lower the laser-cooled mechanical occupation from a value above 1 to a value below 1, thus enabling experiments to probe deeper into the quantum regime. 
Figs~\ref{fig:4}(c) and (d) plot the same quantities as (a) and (b), respectively, but for $\bar{N}=500$. At this higher occupation of the mechanical thermal environment, the relative enhancement in mechanical cooling via zero-photon detection increases.
Utilizing the analysis described above, the steady-state mechanical occupations $\bar{n}_{0}^{(aS)}(t\rightarrow\infty)$ may be solved numerically for a general parameter set. However, for a strongly overcoupled cavity, i.e. $\kappa=\kappa_{ex}$, the analytic result for the ultimate limit of zero-photon-detection-enhanced cooling presented in Ref.~\cite{major2023something} is recovered. 
Beyond the steady-state analysis presented here, we would like to emphasize that for any length string of zero-photon-detection events and non-zero detection efficiency, the mechanical oscillator is cooled below the laser-cooled value.

\hypertarget{link:as_1C}{}\subsubsection{A single-photon detection event}\label{sec:open_as_1c}
A single-photon detection event at time $t$ may be described by setting $\rmd{N}=1$ and $\rmd{t}=0$ in Eq.~\eqref{eq:ajoint_SME_full}.
Here, we again assume that before the single-photon detection event, the optomechanical state is Gaussian, which enables the Isserlis-Wick theorem to be employed. At the instant of the single-photon detection event, the relevant second-order moment transformations are:
\begin{align}
    &\rmi\expval{\ad b}\rightarrow2\rmi\expval{\ad b}+\rmi\dfrac{\expval{a^{\dagger\,2}}\expval{ab}}{\expval{\ad a}}\label{eq:oneclick_as_1},\\
    &\rmi\expval{a \bd}\rightarrow2\rmi\expval{a \bd}+\rmi\dfrac{\expval{a^{2}}\expval{\ad \bd}}{\expval{\ad a}},\label{eq:oneclick_as_2}\\
    &\expval{\ad a}\rightarrow2\expval{\ad a}+\rmi\dfrac{\expval{a^{\dagger\,2}}\expval{a^2}}{\expval{\ad a}},\label{eq:oneclick_as_3}\\
    &\expval{\bd b}\rightarrow\expval{\bd b}+\dfrac{\expval{\ad b}\expval{a\bd}+\expval{ab}\expval{\ad \bd}}{\expval{\ad a}}.\label{eq:oneclick_as_4}
\end{align}
As in the previous section, we show in Appendix~\ref{appendix:zeroclick_as_1} that for the conditions considered here $\expval{a^2}$, $\expval{a^{\dagger\, 2}}$, $\expval{ab}$, and $\expval{\ad\bd}$ are zero for all times, which then implies $\expval{\ad b}=-\expval{a \bd}$ for all times too. Thus, a single-photon detection event is described by the change
\begin{eqnarray}
    \VV_{aS}\rightarrow\VV_{aS}+\mathbf{O}_{aS},\label{eq:matrix_one_click_as}
\end{eqnarray}
where again $\VV_{aS}=\left(2\rmi\expval{\ad b},~\expval{\ad a },~\expval{\bd b}\right)^{\mathrm{T}}$, and $\mathbf{O}_{aS}=\left(2\rmi\expval{\ad b},~\expval{\ad a },~ -\expval{\ad b}^2/\expval{\ad a}\right)^{\mathrm{T}}$.

Notably, a single-photon detection event produces a non-Gaussian optomechanical state. Thus, immediately after a single-photon detection event the Isserlis-Wick theorem is invalid and so Eq.~\eqref{eq:matrix_DE_zero} cannot be used to straightforwardly describe a zero-photon detection event following a single-photon detection event. Instead, one must utilize higher order moments to describe this scenario. Fortunately however, no such approximations were made deriving the laser cooling differential equation and so we may examine the second moments during laser cooling after a single-photon detection event using Eq.~\eqref{eq:laser_cooling_solution_1}. Eq.~\eqref{eq:matrix_DE_zero} may then be used to describe zero-photon detection events a sufficiently long time after the single-photon detection event, when the optomechanical system has returned to a Gaussian state.

When $\kappa_{ex}$ dominates over all other relevant rates, the cavity field may be adiabatically eliminated $(\dot{a}\simeq0)$. In this case, it may be readily shown that $-\expval{\ad b}^2/\expval{\ad a}\simeq\expval{\bd b}$ and so Eq.~\eqref{eq:matrix_one_click_as} becomes $\VV_{aS}\rightarrow 2 \VV_{aS}$. Notably, for an initial thermal mechanical state, this implies that a single-photon detection event leads to a doubling of the mechanical occupation, i.e. $\bar{n}\rightarrow2\bar{n}$.

\subsection{Continuous Stokes interaction}\label{sec:open_s}
If one instead drives the Stokes scattering process, this may be described by setting $H=H_{S}$ in Eq.~\eqref{eq:ajoint_SME_full}, and we consider \hyperlink{link:s_LH}{(i)}~optomechanical parametric amplification, \hyperlink{link:s_ZC}{(ii)}~mechanical cooling via zero-photon detection, and \hyperlink{link:s_1C}{(iii)}~single-photon detection.

\hypertarget{link:s_LH}{}\subsubsection{Optomechanical parametric amplification}\label{sec:open_s_lh}
One may use Eq.~\eqref{eq:ajoint_SME_full} to derive a closed set of coupled differential equations describing the dynamic evolution of the optomechanical system which, for this Stokes process, this closed set of equations is:
\begin{align}
    &\rmi\dfrac{d}{\rmd{t}}\expval{\ad \bd-ab}=-2G\left[\expval{\ad a}+\expval{\bd b}+1\right] \nonumber\\
    &\hspace{90pt}-\rmi\left({\kappa}+\gamma\right)\expval{\ad \bd-ab},\label{eq:LH_s_CDE_1}\\
    &\dfrac{d}{\rmd{t}}\expval{\ad a}=-\rmi G \expval{\ad \bd-ab}-2{\kappa}\expval{\ad a},\label{eq:LH_s_CDE_2}\\
    &\dfrac{d}{\rmd{t}}\expval{\bd b}=-\rmi G \expval{\ad\bd-ab}-2\gamma\expval{\bd b}+2\gamma\bar{N}.\label{eq:LH_s_CDE_3}
\end{align}
One may note that the joint optical-mechanical moments involved in this closed system differ between the anti-Stokes and Stokes scattering processes due to the correlations generated by their respective Hamiltonians and open-system dynamics. The source term $-2G$ (which comes from the $+1$ in the square brackets) is also now present in Eq.~\eqref{eq:LH_s_CDE_1} which is responsible for the parametric amplification commonly associated with Stokes scattering and is not present in Eq.~\eqref{eq:LC_as_CDE_1}. 

As before, these coupled equations without measurement may be written compactly in matrix form as 
\begin{align}
    \dot{\VV}_{S}&=A_{S}\VV_{S}+\mathbf{n}_{S},\label{eq:matrix_eqn_s_1}
\end{align}
where $\VV_{S}=\left(\rmi\expval{\ad \bd-ab},~\expval{\ad a },~\expval{\bd b}\right)^{\mathrm{T}}$, $\mathbf{n}_{S}=\left(-2G,~0,~2\gamma\bar{N}\right)^{\mathrm{T}}$, and
\begin{align}
    A_{S}&=\begin{pmatrix}-\left({\kappa}+\gamma\right) & -2G & -2G\\
    -G & -2{\kappa} & 0\\
    -G & 0 & -2\gamma \end{pmatrix}.\label{eq:Asmatrix}
\end{align}
The steady-state solution of Eq.~\eqref{eq:matrix_eqn_s_1} is
\begin{align}
\VV_{S}(t\rightarrow\infty)&=\begin{pmatrix}-\dfrac{2G\gamma{\kappa}(\bar{N}+1)}{(\gamma+{\kappa})(\gamma{\kappa}-G^2)}\\  \\ \dfrac{G^2\gamma(\bar{N}+1)}{(\gamma+{\kappa})(\gamma{\kappa}-G^2)} \\ \\ \dfrac{\bar{N}\gamma[{\kappa}(\gamma+{\kappa}-G^2)]+G^2{\kappa}}{(\gamma+{\kappa})(\gamma{\kappa}-G^2)}\end{pmatrix},\label{eq:lasercooled_s_long_time}
\end{align}
and the solution with time for the initial condition $\VV_{S}(0)$ may be calculated in the same manner as Eq.~\eqref{eq:laser_cooling_solution_1}. Note that unlike anti-Stokes scattering, the eigenvalues of $A_{S}$ are not negative-definite and the optomechanical fields are amplified exponentially for $G^2\geq\gamma{\kappa}$ as well as displaying a dynamic instability~\cite{qian2012quantum,cryer2023second}. Henceforth, we will restrict ourselves to the regime where $G^2<\gamma{\kappa}$ when the steady-state solution in Eq.~\eqref{eq:lasercooled_s_long_time} is valid.

\hypertarget{link:s_ZC}{}\subsubsection{Stokes interaction and continuous-time zero-photon detection}\label{sec:open_s_zc}
Considering a string of zero-photon detection events, measured with efficiency $0<{\eta}\leq1$, and beginning with the optomechanical system in Gaussian states, we again make use of Isserlis-Wick theorem and evaluate Eq.~\eqref{eq:ajoint_SME_full} with $dN=0$ to obtain
\begin{widetext}
\begin{align}
    &\rmi\dfrac{d}{\rmd{t}}\expval{\ad \bd}=-G\left[\expval{\ad a}+\expval{\bd b} + 1\right]-\rmi\left({\kappa}+\gamma\right)\expval{\ad \bd}-2\rmi{\eta}\kappa_{ex}\expval{\ad a}\expval{\ad \bd}-2\rmi{\eta}\kappa_{ex}\expval{a^{\dagger\,2}}\expval{a \bd},\label{eq:zc_cde_1}\\
    &\rmi\dfrac{d}{\rmd{t}}\expval{ab}=G\left[\expval{\ad a}+\expval{\bd b} + 1\right]-\rmi\left({\kappa}+\gamma\right)\expval{ab}-2\rmi{\eta}\kappa_{ex}\expval{\ad a}\expval{a b}-2\rmi{\eta}\kappa_{ex}\expval{a^2}\expval{\ad b},\label{eq:zc_cde_2}\\
    &\dfrac{d}{\rmd{t}}\expval{\ad a}=-\rmi G \expval{\ad \bd-ab}-2{\kappa}\expval{\ad a}-2{\eta}\kappa_{ex}\expval{\ad a}^2-2{\eta}\kappa_{ex}\expval{a}^2\expval{a^{\dagger\,2}},\label{eq:zc_cde_3}\\
    &\dfrac{d}{\rmd{t}}\expval{\bd b}=-\rmi G \expval{\ad \bd-ab}-2\gamma\expval{\bd b}+2\gamma\bar{N}-2{\eta}\kappa_{ex}\expval{a\bd}\expval{\ad b}-2{\eta}\kappa_{ex}\expval{\ad\bd}\expval{a b}.\label{eq:zc_cde_4}
\end{align}
\end{widetext}
Using the same reasoning as for the anti-Stokes case, the moments describing correlations not generated by the interplay of unitary dynamics and Lindbladian dissipation, such as $\expval{a^2}$,  $\expval{a^{\dagger\, 2}}$, $\expval{a\bd}$, and $\expval{\ad b}$, remain zero for states where they are zero at $t=0$. As this is true for the initial states discussed here, we may neglect these terms and again obtain a matrix differential equation including continuous zero-photon detection given by
\begin{eqnarray}
    \dot{\VV}_{S}&=&A_{S}\VV_{S}+\mathbf{n}_{S}-{\eta}\kappa_{ex}\mathbf{Z}_{S},\label{eq:matrix_DE_zero_s}
\end{eqnarray}
where $\mathbf{Z}_{S}=\left(4\rmi\expval{\ad \bd}\expval{\ad a},~2\expval{\ad a}^2,~-2\expval{\ad \bd}^2\right)^\mathrm{T}$, $A_{S}$ and $\mathbf{n}_{S}$ are defined in and above Eq.~\eqref{eq:Asmatrix}, and we redefine $ \VV_{S}=\left(2\rmi\expval{\ad \bd},~\expval{\ad a },~\expval{\bd b}\right)^{\mathrm{T}}$. It should be noted that, as in Section \ref{sec:no_open_s}, a lower bound $\eta^{*}$ is found for the efficiency required for mechanical cooling via zero-photon detection while driving the Stokes interaction to cool the mechanical state below its initial occupation. In the presence of open-system dynamics, $\eta^{*}$ is calculated numerically.

\begin{figure*}
    \centering
    \includegraphics[width=0.8\textwidth]{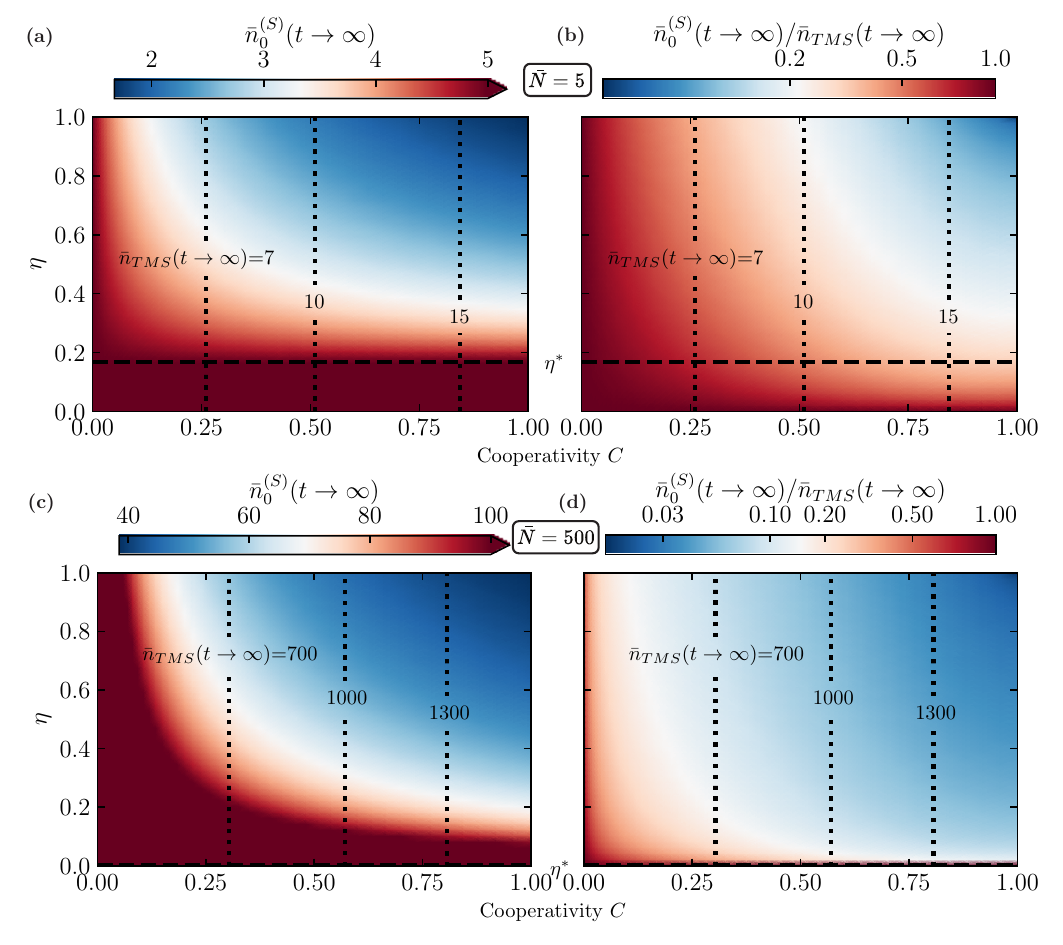}
    \caption{Mechanical cooling via the Stokes interaction and continuous zero-photon detection. (a) The mechanical occupation after continual zero-photon detection $\bar{n}_{0}^{(S)}(t\rightarrow\infty)$ as a function of detection efficiency $\eta$ and optomechanical cooperativity $C$ for $\bar{N}=5$. To emphasize the effect of cooling via zero-photon detection, all occupations above $\bar{N}$ are shown as a single colour (dark red). (b) The ratio between the steady-state mechanical occupations after continual zero-photon detection and unconditioned optomechanical parametric amplification $\bar{n}_{0}^{(S)}(t\rightarrow\infty)/\bar{n}_{TMS}(t\rightarrow\infty)$. The horizontal dashed line indicates the efficiency ${\eta}^*$, above which the mechanical mode is cooled via zero-photon detection when the Stokes process is driven. For the simulation parameters, this efficiency is ${\eta}^*\approx0.17$. Plots (c) and (d) illustrate the same quantities as (a) and (b) respectively but for $\bar{N}=500$. Note that here ${\eta}^*\gtrsim0$ and zero-photon detection with any finite efficiency results in cooling and all occupations above 100 are again shown as a single colour (dark red). The vertical dotted lines indicate the unconditioned steady-state mechanical occupations when the Stokes process is continually driven $\bar{n}_{TMS}(t\rightarrow\infty)$. The values of the other parameters used in the simulation are ${\kappa}=\kappa_{ex}=40$ and $\gamma=1$. All rates are normalized by $\gamma$.}
    \label{fig:5}
\end{figure*}

In Figs~\ref{fig:5}(a) and (c), we plot the steady-state solution of Eq.~\eqref{eq:matrix_DE_zero_s} for the mechanical occupation $\bar{n}_{0}^{(S)}(t\rightarrow\infty)$ as a function of the detection efficiency $\eta$ and optomechanical cooperativity $C=G^2/({\kappa}\gamma)$ for $\bar{N}=5$ and $\bar{N}=500$, respectively. And, in Figs~\ref{fig:5}(b) and (d) we plot these quantities divided by the unconditioned steady-state mechanical occupation realized via continuous driving of the Stokes interaction $\bar{n}_{TMS}(t\rightarrow\infty)$. As in Fig.~\ref{fig:4}, the vertical lines indicate unconditioned steady-state mechanical occupations $\bar{n}_{TMS}(t\rightarrow\infty)$, which in this case are all greater than the occupation of the mechanical bath due to the heating effect of the Stokes interaction. However, for any efficiency above the threshold efficiency ${\eta}^*$, continuous zero-photon detection can used to cool the mechanical occupation below the occupation of the bath. For the parameters chosen in Figs~\ref{fig:5}(a) and (b), the threshold efficiency is given by ${\eta}^*\approx0.17$ for $\bar{N}=5$. Meanwhile in Figs~\ref{fig:5}(c) and (d), the higher occupation of the mechanical environment of $\bar{N}=500$ produces a threshold efficiency ${\eta}^*\gtrsim0$. 

\begin{figure*}
    \centering
    \includegraphics[width=0.8\textwidth]{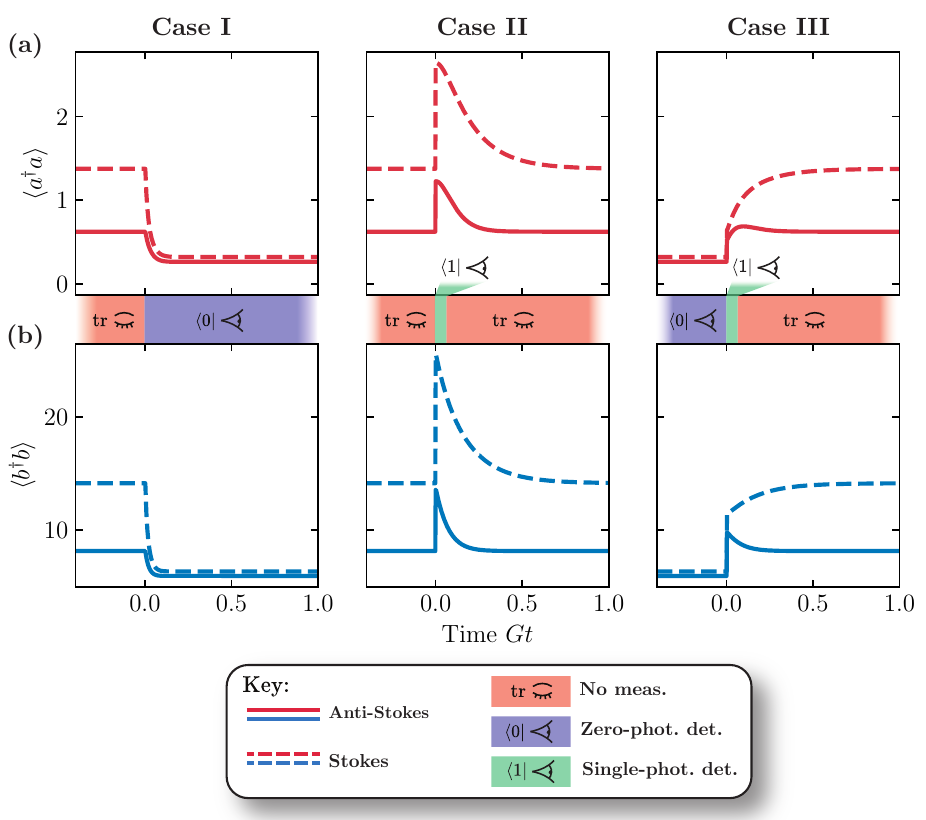}
    \caption{Comparison between continuous mechanical cooling via zero-photon detection utilizing the anti-Stokes (solid lines)  or Stokes interaction (dashed lines) for three cases. 
    In Case I, a period of continuous laser cooling/parametric amplification is followed by continuous zero-photon detection. In Case II, a single-photon detection occurs in the middle of periods of unconditioned evolution, illustrating a transient excursion from equilibrium. In Case III, the quasi-equilibrium reached after a continuous period of zero-photon detection is followed by a single-photon detection and then an unmonitored evolution. (a) The effect of each case on the expectation value of the optical number operator $\expval{\ad a}$. (b) The effect of the same measurement records on the mechanical occupation $\expval{\bd b}$. The measurements made at each point in time are indicated by the colour of the bar between (a) and (b), detailed in the key. The values of simulation parameters are $\bar{N}=10$, ${\kappa}=\kappa_{ex}=3$, $\gamma=1$, $G=1$ and ${\eta}=1$. All rates are normalized by $G$.}
    \label{fig:6}
\end{figure*}

\hypertarget{link:s_1C}{}\subsubsection{A single-photon detection event}\label{sec:open_s_1c}
Finally, the change in expectation values of the moments considered previously can be calculated at the time of a single-photon detection event.
For the zero-mean Gaussian states considered here, the Isserlis-Wick theorem is valid and this change is given by
\begin{align}
    &\rmi\expval{\ad \bd}\rightarrow2\rmi\expval{\ad \bd}+\rmi\dfrac{\expval{a^{\dagger\,2}}\expval{a\bd}}{\expval{\ad a}}\label{eq:oneclick_s_1},\\
    &\rmi\expval{ab}\rightarrow2\rmi\expval{a b}+\rmi\dfrac{\expval{a^{2}}\expval{\ad b}}{\expval{\ad a}},\label{eq:oneclick_s_2}\\
    &\expval{\ad a}\rightarrow2\expval{\ad a}+\rmi\dfrac{\expval{a^{\dagger\,2}}\expval{a^2}}{\expval{\ad a}},\label{eq:oneclick_s_3}\\
    &\expval{\bd b}\rightarrow\expval{\bd b}+\dfrac{\expval{\ad b}\expval{a\bd}+\expval{ab}\expval{\ad \bd}}{\expval{\ad a}}.\label{eq:oneclick_s_4}
\end{align}
We again consider only those moments that are non-zero (see Appendix \ref{appendix:zeroclick_as_1}), giving the Stokes analogue of Eq.~\eqref{eq:matrix_one_click_as} as
\begin{eqnarray}
    \VV_{S}\rightarrow\VV_{S}+\mathbf{O}_{S},\label{eq:matrix_one_click_s}
\end{eqnarray}
where $\mathbf{O}_{S}=\left(2\rmi\expval{\ad \bd},~\expval{\ad a },~ -\expval{\ad \bd}^2/\expval{\ad a}\right)^{\mathrm{T}}$. 

Similarly to the anti-Stokes case, immediately after a single-photon detection event, the optomechanical state is non-Gaussian. Thus Eq.~\eqref{eq:matrix_DE_zero_s}, which assumes a Gaussian optomechanical state, cannot be used at this time. However, one may use Eq.~\eqref{eq:matrix_eqn_s_1} immediately after a single-photon detection event, which does not assume the optomechanical state is Gaussian. Furthermore,in the adiabatic regime, Eq.~\eqref{eq:matrix_one_click_s} becomes $\VV_{S}\rightarrow 2 \VV_{S}+\left(0,~0,~1\right)^{\mathrm{T}}$. Hence, for an initial mechanical thermal state, we recover that a single-photon detection event changes the mean occupation according to $\bar{n}\rightarrow 2\bar{n}+1$.

\subsection{Comparison between the anti-Stokes and Stokes processes}
In Fig.~\ref{fig:6} we compare the results of Sections~\ref{sec:open_as} and \ref{sec:open_s}. To ensure this comparison between the anti-Stokes and Stokes scattering processes is fair and to highlight the relevant effects of photon detection on the mechanical and optical populations, we consider three experimentally relevant cases in Figs~\ref{fig:6}(a) and (b) comprising periods of unconditional evolution, continuous zero-photon detection and single-photon detection events. 

In Case I, the laser-cooled/parametrically-amplified steady state is reached after a period of unconditioned evolution and then a continuous string of zero-photon detection events is recorded. The unconditioned steady state illustrates the usual features of the anti-Stokes and Stokes processes: the former drives a state swap between the mechanical and optical fields, laser cooling the mechanical mode. 
The latter generates excitations in the mechanical and optical modes, heating both until an equilibrium is reached where loss rates balance optomechanical gain. Observing a continuous string of zero-photon detection events cools the mechanical state for the anti-Stokes interaction and can, surprisingly, cool for the Stokes interaction too. For this cooling to exceed heating while driving the Stokes interaction, the measurement efficiency must exceed $\eta^{*}$.
It is also worth noting that while zero-photon detection during the anti-Stokes interaction always results in lower optomechanical occupations than the Stokes interaction, as no mechanical excitations are created during anti-Stokes scattering, the relative cooling of the state is larger for zero-photon detection while driving the Stokes interaction due to the ``surprise'' factor during the gain process. This difference in the relative enhancement of cooling by zero-photon detection may also be seen by comparing Figs~\ref{fig:4}(b) and \ref{fig:5}(b).   

In Case II, the detection of a single photon during unconditioned evolution results in an increase in the occupation of the optical cavity and mechanical modes for both the anti-Stokes and Stokes interactions. 
In the limit of infinitesimal measurement times used to derive the SME given in Eq.~\eqref{eq:SME}, the optical occupation doubles for both the anti-Stokes and Stokes interactions [cf. Eqs~\eqref{eq:matrix_one_click_as} and \eqref{eq:matrix_one_click_s}]. However, outside the adiabatic regime, the mechanical occupation increases by a factor less than two due to the interplay of the system decay rates. 

Finally, in Case III, a continuous string of zero-photon detection events is followed by a single-photon detection and then a period of unconditioned evolution. For the simulation parameters used in Fig.~~\ref{fig:6}, the optical and mechanical populations illustrate a brief excursion above the steady-state values reached via continuous zero-photon detection, before then returning to their laser-cooled/parametrically-amplified steady state.

\section{Conclusions and Outlook}
Here, we have provided a theoretical description of mechanical cooling utilizing the anti-Stokes or Stokes interactions and zero-photon detection. Our analysis describes pulsed as well as continuously monitored cavity optomechanical interactions incorporating detection inefficiencies and mechanical open-system dynamics. For both of these scenarios, we have shown that mechanical cooling via zero-photon detection is enhanced for the anti-Stokes interaction and, surprisingly, enabled for the Stokes interaction. The results of Section~\ref{sec:SME_non_adiabatic} detail the anti-Stokes interaction, which was experimentally explored in Ref.~\cite{major2023something} to enhance mechanical cooling via zero-photon detection beyond the limits of laser cooling. Importantly, the enhanced cooling is realized for any non-zero detection efficiency and thus the techniques introduced here can be readily utilized across a wide range of current and future optomechanics experiments. For the Stokes interaction, there is a competition between optomechanical parametric amplification and cooling from zero-photon detection and we have determined the detection efficiency required for the cooling to overcome the heating process. 

Beyond zero-photon detection, the analysis presented here also highlights contributions beyond first order in $G$ to the operation to the mechanical mode following a pulsed optomechanical interaction and $n$-photon detection. Namely, $n$-photon detection does not directly correspond to $n$-phonon addition or subtraction for the Stokes or anti-Stokes scattering processes, respectively [cf. Eqs~\eqref{eq:rho_as_n_occupation} and \eqref{eq:rho_s_n_occupation}]. Interestingly, these contributions are shown to be different for the Stokes and anti-Stokes scattering processes, which could be utilized in future works including thermometry~\cite{mehboudi2019thermometry}, quantum thermodynamics~\cite{vinjanampathy2016quantum}, and studies of open quantum systems~\cite{breuer2002theory}.

The zero-photon-detection-based techniques we have introduced, provide new routes to prepare mechanical oscillators in higher-purity states. Building on the growing interest in optomechanics utilizing photon counting~\cite{vanner2013quantum, borkje2011proposal, marshall2003towards, cohen2015phonon, Enzian2021, Enzian2021_2, Patel2021, patil2022measuring, lee2012macroscopic, riedinger2016non, Milburn2016, ringbauer2018generation, howard2019quantum, clarke2019growing, kanari2022two, cryer2023second, galinskiy2023non}, these zero-photon measurement techniques may be readily employed in a range of physical systems. In particular, when the Stokes process is driven for single-phonon-addition operations~\cite{vanner2013quantum} for nonclassical state preparation, continuous zero-photon detection can be utilized prior to single-photon detection to both avoid the mechanical heating via parametric amplification and to reduce the initial thermal occupation. This can allow greater non-classicality to be generated without the need for a second anti-Stokes interaction.

\begin{acknowledgments}
We acknowledge useful discussions with Lewis A. Clark, Rufus Clarke, Artie Clayton-Major,  Lars Freisem, Daniel Hodgson, Gerard J. Milburn, and John J. Price. This project was supported by UK Research and Innovation (MR/S032924/1, MR/X024105/1), the Engineering and Physical Sciences Research Council (EP/T031271/1, EP/P510257/1), the Science and Technology Facilities Council (ST/W006553/1), the Royal Society, and the Aker Scholarship.
\end{acknowledgments}

\appendix

\section{Simplifying the dynamics}\label{appendix:zeroclick_as_1}

Here, we provide further details for the simplifications made to the equations for the second-order moments in Section~\ref{sec:SME_non_adiabatic}. In particular, we focus on the equations describing laser cooling, cooling via zero-photon detection, and single-photon detection events realized via the anti-Stokes interaction and detail why $\expval{a^2}$, $\expval{a b}$, $\expval{b^2}$, and their Hermitian conjugates are zero for all times. Notably, analogous arguments apply for the equations describing optomechanical parametric amplification, cooling via zero-photon detection, and single-photon detection events realized via the Stokes interaction, which in this case imply $\expval{a^2}$, $\expval{a \bd}$, $\expval{b^2}$, and their Hermitian conjugates are zero for all times. 

Before discussing the system dynamics, we will first note some relevant properties of the assumed initial optomechanical state. Here, we assume an initially phase-symmetric and separable optical-mechanical state given by $\ket{0}_{l}\prescript{}{l}{\bra{0}}\otimes\rho_{\bar{n}}$. This initial state and all phase-symmetric states have reduced density operators that are diagonal in the number basis and, as a consequence:
\begin{align}
&\expval{a}=\expval{\ad}=\expval{b}=\expval{\bd}=0,\label{eq:phase_sym_1}\\
&\expval{a^2}=\expval{b^2}=\expval{a^{\dagger\,2}}=\expval{b^{\dagger\,2}}=0.\label{eq:phase_sym_2}
\end{align}
Moreover, for a separable optical-mechanical state, Eq.~\eqref{eq:phase_sym_1} implies
\begin{eqnarray}
    \expval{ab}=\expval{a \bd}=\expval{\ad b}=\expval{\ad \bd}=0.\label{eq:separable_1}
\end{eqnarray}
Notably, Eqs~\eqref{eq:SME} and \eqref{eq:ajoint_SME_full} preserve the phase symmetry of an initially phase-symmetric optomechanical state.

\subsection{Laser cooling and cooling via zero-photon detection}
Eqs~\eqref{eq:zc_as_cde_1} to \eqref{eq:zc_as_cde_4} describe laser cooling (${\eta}=0$) and cooling via zero-photon detection ($0<{\eta}\leq1$) for the moments $\expval{\ad b}$, $\expval{a \bd}$, $\expval{\ad a}$, and $\expval{\bd b}$. However, the moments in Eqs~\eqref{eq:zc_as_cde_1} and \eqref{eq:zc_as_cde_2} are Hermitian conjugates of one other and thus Eqs~\eqref{eq:zc_as_cde_1}, \eqref{eq:zc_as_cde_3}, and \eqref{eq:zc_as_cde_4} describe the evolution of the three unique moments $\expval{\ad b}$, $\expval{\ad a}$, and $\expval{\bd b}$.

The three remaining unique second-order moments describing laser cooling and cooling via zero-photon detection are $\expval{a^2}$, $\expval{a b}$, and $\expval{b^2}$, which evolve according to
\begin{widetext}
\begin{align}
    &\dfrac{d}{\rmd{t}}\expval{a^2}=-2\rmi G \expval{ab}-2{\kappa}\expval{a^2}-4{\eta}\kappa_{ex}\expval{\ad a}\expval{a^2},\label{eq:zc_as_cde_5}\\
    &\dfrac{d}{\rmd{t}}\expval{a b}=-\rmi G\left[\expval{a^2}+\expval{b^2}\right]-\left({\kappa}+\gamma\right)\expval{a b}-2{\eta}\kappa_{ex}\left[\expval{ab}\expval{\ad a}+\expval{a^2}\expval{\ad b}\right],\label{eq:zc_as_cde_6}\\
    &\dfrac{d}{\rmd{t}}\expval{b^2}=-2\rmi G \expval{ab}-2\gamma\expval{b^2}-4{\eta}\kappa_{ex}\expval{\ad b}\expval{a b}.\label{eq:zc_as_cde_7}
\end{align}
\end{widetext}
Here, the Isserlis-Wick theorem has been used to simplify the terms proportional to ${\eta}\kappa_{ex}$, which originate from zero-photon measurements. By inspecting Eqs~\eqref{eq:zc_as_cde_5}, \eqref{eq:zc_as_cde_6}, and \eqref{eq:zc_as_cde_7}, it is clear that if at time $t=0$, $\expval{a^2}=\expval{a b}=\expval{b^2}=0$, then $\expval{a^2}=\expval{a b}=\expval{b^2}=0$ for all time too. Note that the equation describing the evolution of $\expval{\ad\bd}$ is given by the Hermitian conjugate of Eq.~\eqref{eq:zc_as_cde_6}.

\subsection{A single-photon detection event}
Together, Eqs~\eqref{eq:oneclick_as_1} to \eqref{eq:oneclick_as_4} describe how the three unique moments $\expval{\ad b}$, $\expval{\ad a}$, and $\expval{\bd b}$ change due to a single-photon detection event. As in the previous section, the three remaining unique second-order moments are $\expval{a^2}$, $\expval{a b}$, and $\expval{b^2}$, which transform according to
\begin{align}
    &\expval{a^2}\rightarrow3\expval{a^2},\label{eq:oneclick_as_5}\\
    &\expval{a b}\rightarrow2\expval{ab}+\dfrac{\expval{\ad b}\expval{a^2}}{\expval{\ad a}},\label{eq:oneclick_as_6}\\
    &\expval{b^2}\rightarrow\expval{b^2}+\dfrac{2\expval{\ad b}\expval{a b}}{\expval{\ad a}},\label{eq:oneclick_as_7}
\end{align}
during a single-photon detection event. Here, we have assumed that the optomechanical state at time $t$ is Gaussian such that the Isserlis-Wick theorem can used to simplify the right-hand side of Eqs~\eqref{eq:oneclick_as_5}, \eqref{eq:oneclick_as_6}, and \eqref{eq:oneclick_as_7}.  By inspecting these equations, one may observe that if before the single-photon detection, $\expval{a^2}=\expval{a b}=\expval{b^2}=0$, then after the single-photon detection $\expval{a^2}=\expval{a b}=\expval{b^2}=0$ too. Note again that the equation describing the evolution of $\expval{\ad\bd}$ is given by the Hermitian conjugate of Eq.~\eqref{eq:oneclick_as_6}.

\providecommand{\noopsort}[1]{}\providecommand{\singleletter}[1]{#1}%

\end{document}